\documentclass[prb,twocolumn,showpacs,preprintnumbers,
amsmath,amssymb,superscriptaddress]{revtex4}

\usepackage{graphicx}
\usepackage{dcolumn}
\usepackage{bm}
\newcommand{\beq}{\begin{equation}}
\newcommand{\eeq}{\end{equation}}
\newcommand{\beqa}{\begin{eqnarray}}
\newcommand{\eeqa}{\end{eqnarray}}

\newcommand{\w}{\omega}

\newcommand{\ket}[1]{\left| #1 \right\rangle}
\newcommand{\bra}[1]{\left\langle #1 \right|}

\newcommand{\Xp}{x_{+}}
\newcommand{\Xm}{x_{-}}
\begin{document}

\preprint{V210105}

\title{Quantum Computing with Spin Qubits Interacting Through Delocalized Excitons: Overcoming Hole Mixing}
\author{Brendon~W.~Lovett}
\email{brendon.lovett@materials.oxford.ac.uk}
\affiliation{Department of Materials, Oxford University, 
Oxford OX1 3PH, United Kingdom}
\author{Ahsan~Nazir}
\affiliation{Department of Materials, Oxford University, 
Oxford OX1 3PH, United Kingdom}
\author{Ehoud~Pazy}
\affiliation{Chemistry Department,
 Ben - Gurion University of the Negev, P.O. Box 653, Beer-Sheva 84105, Israel}
\author{Sean~D.~Barrett}
\affiliation{Hewlett-Packard Laboratories, Filton Road,
Stoke Gifford, Bristol BS34 8QZ, United Kingdom}
\author{Timothy~P.~Spiller}
\affiliation{Hewlett-Packard Laboratories, Filton Road,
Stoke Gifford, Bristol BS34 8QZ, United Kingdom}
\author{G.~Andrew~D.~Briggs}
\affiliation{Department of Materials, 
Oxford University, Oxford OX1 3PH, United Kingdom}

\date{\today}
\begin{abstract}
As a candidate scheme for controllably coupled qubits, we consider two quantum dots, each doped with a single electron. The spin of the electron defines our qubit basis and trion states can be 
created by using polarized light; we show that the form of 
the excited trion depends on the state of the qubit. 
By using the Luttinger-Kohn
Hamiltonian we calculate the form of these trion states in the presence 
of light-heavy hole mixing, and show that they can interact through 
both the F\"orster transfer and static dipole-dipole interactions. 
Finally, we demonstrate that by using chirped laser pulses, it 
is possible to perform a two-qubit gate in this system by 
adiabatically following the eigenstates as a function of 
laser detuning. These gates are robust in that they operate 
with any realistic degree of hole mixing, and for either 
type of trion-trion coupling.
\end{abstract}

\maketitle

\section{\label{intro}Introduction}

In the quest for a solid state quantum information processor, there is great attraction in combining the relatively long coherence times of spins with the speed and versatility of optical manipulation. There
have been many papers describing different 
ways to embody a qubit by using the two levels of a confined 
spin-1/2 electron.~\cite{loss98, sham02, kane98,imamoglu00,spinqu,troiani03} 
The direct interaction between two such spin qubits is often quite weak, but they can be enhanced by exploiting degrees of freedom which lie outside the computation Hilbert space. For example, spin information can be transferred to spatial degrees of freedom in a double quantum dot structure,~\cite{troiani03} or to photons in an electromagnetic cavity.~\cite{imamoglu99}
Recent proposals~\cite{pazy03, calarco03, nazir04} have 
described ways in which spins might be coupled by using 
polarized light to selectively create trion states. Such spin selective
charge excitations benefit directly from the recent progress in 
ultrafast optoelectronics: both the coherent manipulation 
of excitons in quantum dots (QDs)~\cite{Rabi,Bonade98,Li03} 
and spin selective optical transitions~\cite{Yokoi} have been demonstrated. Moreover, once a 
gate operation is complete, it is possible to arrange that all 
population returns to the qubit subspace, and the quantum 
device therefore benefits from the robust coherence 
properties~\cite{hanson03, kroutvar04, golovach04} of the electron spin.

In Ref.~\onlinecite{nazir04}, we demonstrated that an entangling CPHASE 
gate can be performed between single spins on each of two adjacent QDs by the
spin selective excitation of a single, delocalized, exciton. The required 
delocalization occurs when the dots are (near) resonant, and if 
they interact through the F\"orster energy transfer 
mechanism.~\cite{forster48, dexter53} The proposal is based on the Pauli blocking mechanism~\cite{Warburton97} and is valid only
in the special case of {\it no light-heavy hole mixing}, 
which, though sometimes applicable,~\cite{bayer99} is not generally true for most QD systems.~\cite{sheng03}  
We shall here demonstrate an alternative 
method of performing a two-qubit gate, which 
solves this problem. Following Ref.~\onlinecite{calarco03}, it is based on using chirped laser pulses to perform adiabatic quantum gate operations.

To develop this model, the following steps are necessary. We shall first describe 
how the QDs employed in our quantum information
implementation scheme are modeled.
We describe the hole sub-band mixing in terms of a four 
band Luttinger-Kohn model,~\cite{luttinger} and show how this affects the 
coupling of confined charge carriers to a laser field (Section~\ref{QD}).
We then consider a coupled QD system
and present the dependence of the F\"orster transfer operator on the 
angular momenta of the excitonic states which it connects. We shall
derive the form of the F\"orster transfer interaction for two coupled trions and hence 
write down an effective Hamiltonian for two interacting QDs coupled to a laser field,
in the presence of hole mixing (Section~\ref{coupledQD}). 
The proposal for performing quantum gates by creating trions adiabatically is then discussed (Section~\ref{gates}), and 
we describe two different modes in which the two 
qubit gate can be operated. We shall demonstrate 
that the adiabatic scheme circumvents the hole mixing problem
and briefly discuss why it should also reduce 
phonon decoherence. State measurement, preparation and scalability will be discussed next (Section~\ref{meas}) and then we summarize (Section~\ref{summary}).

\section{Single Quantum Dot Model}
\label{QD}

Let us consider self-assembled QDs with strong confinement along the 
growth direction $z$, which is also the QD symmetry axis.
This type of QD can be produced in materials such as InGaAs by using the Stranski-Krastanow method,~\cite{SK} 
which may allow the realization of a controllably coupled many dot system.
Such QDs exist in the
strong confinement regime, in which the typical size, $L$, of the QD
in the growth direction is of the order of $10-20$~nm.
In this regime, the Coulomb interaction between charge carriers scales
as $1/L$, but the single-particle excitation energy has a $1/L^2$ dependence. Excitonic wavefunctions can therefore be modelled by products of single particle electron and hole states, with Coulomb effects being introduced by using first order perturbation theory. This approximation results in a shift to the excitonic energy but does not lead to the entanglement of electron and hole.
The effective mass and envelope function
approximations reveal that the electronic states inside the QD exhibit
atomic-like symmetries, which have been identified experimentally.~\cite{atomlike}

The wavefunction for a single particle in a QD can be described by a product of a {\it Bloch function} $U$, which has the periodicity of the atomic lattice, and an {\it envelope function} $\phi$, which describes the amplitude modulation of the wavefunction that is imposed by the confinement potential.
Henceforth we shall only consider the lowest energy envelope function 
for both the conduction and valence bands (which has no nodes in both cases), 
and neglect any mixing with higher envelopes. This approximation is 
discussed in Ref.~\onlinecite{tsitsishvili98}: we use it for clarity and our scheme does not depend on it; a more thorough desciption of the electronic structure of self-assembled QDs is presented in Ref.~\onlinecite{Williamson}.

The eigenstates of the 
angular momentum operators, $\hat{J}$ and $\hat{J_z}$ for the six hole states 
closest to the top of the valence band can be represented by:~\cite{basu97}
\beqa
\label{bloch1}
\ket{3/2_h, 3/2} &=& \frac{f_{hh}(r)}{\sqrt{2}}\ket{(X+iY)\alpha},\\
\ket{3/2_h, -3/2} &=& \frac{f_{hh}(r)}{\sqrt{2}}\ket{(X-iY)\beta},\\
\ket{3/2_h, 1/2} &=& \frac{f_{lh}(r)}{\sqrt{6}}\left[\ket{(X+iY)\beta} 
- \ket{2Z\alpha}\right],\\
\ket{3/2_h, -1/2} &=& \frac{f_{lh}(r)}{\sqrt{6}}\left[\ket{(X-iY)\alpha} 
+ \ket{2Z\beta}\right],
\label{bloch4}\\
\ket{1/2_h, 1/2} &=& -\frac{f_{so}(r)}{\sqrt{3}}\left[\ket{(X+iY)\beta} 
+ \ket{Z\alpha}\right],\\
\ket{1/2_h, -1/2} &=& -\frac{f_{so}(r)}{\sqrt{3}}\left[\ket{(X-iY)\alpha} 
- \ket{Z\beta}\right].
\label{bloch6}
\eeqa
We have labeled the Bloch functions $U$ by using the notation 
$\ket{J_h, J_z}$. The first two states correspond to heavy holes ($hh$); 
the next two are light holes ($lh$) and the last two are split-off holes 
($so$). The functions 
$f_i$ describe the radial dependence of each Bloch function type 
$i \in \{hh, lh, so\}$;
 $\alpha$ and $\beta$ are the up and down spin states respectively. 
The $X$, $Y$, and $Z$ represent orbital wavefunctions as follows:
\beqa
\bra{\bf r}X\rangle&=& \sqrt{\frac{3}{4\pi}}\sin\theta\cos\phi\\
\bra{\bf r}Y\rangle&=& \sqrt{\frac{3}{4\pi}}\sin\theta\sin\phi\\
\bra{\bf r}Z\rangle&=& \sqrt{\frac{3}{4\pi}}\cos\theta.
\eeqa
The electron states are simply:
\beqa
\label{bloch7}
\ket{1/2_e, 1/2} &=&g(r)\ket{S\alpha},\\
\ket{1/2_e, -1/2} &=&g(r)\ket{S\beta}.
\label{bloch8}
\eeqa
$\bra{\bf r}S\rangle = 1/\sqrt{4\pi}$ is the isotropic orbital function 
and $g(r)$ is the radial dependence of the electron's wavefunction.

We now note that:
\beqa
\int \bra{\bf r}X\rangle x  \bra{\bf r}S\rangle d\Omega &=& 
\int \bra{\bf r}Y\rangle y  \bra{\bf r}S\rangle d\Omega =\nonumber\\ 
\int \bra{\bf r}Z\rangle z  \bra{\bf r}S\rangle d\Omega &=& 
\frac{r}{\sqrt{3}}
\label{rel1}
\eeqa
where $d\Omega$ is the infinitesimal solid angle. We also see that:
\beqa
\int \bra{\bf r}X\rangle y  \bra{\bf r}S\rangle d\Omega &=& 
\int \bra{\bf r}X\rangle z  \bra{\bf r}S\rangle d\Omega =0\\ 
\int \bra{\bf r}Y\rangle x  \bra{\bf r}S\rangle d\Omega &=& 
\int \bra{\bf r}Y\rangle z  \bra{\bf r}S\rangle d\Omega =0\\
\int \bra{\bf r}Z\rangle x  \bra{\bf r}S\rangle d\Omega &=& 
\int \bra{\bf r}Z\rangle y  \bra{\bf r}S\rangle d\Omega = 0.
\label{rel2}
\eeqa
These relations will be important in the following discussion.

\subsection{Hole Mixing}
\label{mixing}

Most semiconductors exhibit mixing
of the heavy and light hole sub-bands, which we shall now describe
by using the Luttinger-Kohn model.~\cite{luttinger}
The electron eigenstates of bulk 
semiconductors may be characterized by the crystal momentum 
wave-vector, ${\bf k} = \{k_x, k_y, k_z\}$. The coupling between 
light and heavy holes is described by a four band 
Luttinger-Kohn Hamiltonian,~\cite{luttinger} so long as the 
split off holes are energetically distant enough that coupling to this band can be neglected.~\cite{basu97} In the basis $\{\ket{J_z=+3/2}, 
\ket{J_z=+1/2}, \ket{J_z=-1/2}, \ket{J_z=-3/2}\}$, the Hamiltonian is 
written:
\beq
\label{luttinger}
{\cal H}= \left( \begin{array}{cccc}
H_{hh} & -b &-c & 0 \\
-b^\ast & H_{lh} & 0 & -c \\  
-c^\ast & 0 & H_{lh} & b\\   
0 & -c^\ast & b^\ast & H_{hh} \\  \end{array} \right).
\eeq
The uncoupled heavy hole Hamiltonian, $H_{hh}$ is:
\beq
H_{hh} = \frac{\hbar^2k_z^2}{2m_0}(\gamma_1-2\gamma_2) 
+ \frac{\hbar^2(k_x^2+k_y^2)}{2m_0}(\gamma_1+\gamma_2).
\eeq
The uncoupled light hole Hamiltonian, $H_{lh}$ is:
\beq
H_{ll} = \frac{\hbar^2k_z^2}{2m_0}(\gamma_1+2\gamma_2) 
+ \frac{\hbar^2(k_x^2+k_y^2)}{2m_0}(\gamma_1-\gamma_2).
\eeq
The mixing parameters are:
\beq
c = \frac{\sqrt{3}\hbar^2}{2m_0}[\gamma_2(k_x^2-k_y^2)-2i\gamma_3k_xk_y]
\eeq
and 
\beq
\label{b}
b=\frac{\sqrt{3}\hbar^2}{m_0}\gamma_3k_z(k_x-ik_y).
\eeq
$\gamma_1$, $\gamma_2$ and $\gamma_3$ are the Luttinger 
parameters~\cite{basu97, haug90} and $m_0$ is the free electron mass.

For the quantum confined states that are found in semiconductor 
nanostructures, the crystal momentum is no longer a good quantum number. 
It is therefore necessary to replace each component of ${\bf k}$ with its 
expectation value, taken over the hole envelope function, $\phi(\bf r)$.~\cite{basu97} 
That is:
\beq
{\bf k} \rightarrow -i\int \phi({\bf r}) \nabla \phi({\bf r}) d{\bf r} 
\equiv \langle {\bf k}\rangle.
\eeq
If the envelope function has a well-defined parity then $\langle k_x 
\rangle = \langle k_y\rangle = \langle k_z \rangle = 0$, and so, 
from Eq.~\ref{b}, $b=0$. Therefore, the Hamiltonian, Eq.~\ref{luttinger} 
decouples and acts in two separate two-dimensional Hilbert spaces. 
We find that the hole eigenstates are given by the two pairs:
\beqa
\ket{h_+} &=& \sqrt{1-\epsilon^2} \ket{J_z = +3/2} + \epsilon 
\ket{J_z = -1/2},\nonumber\\
\ket{h^\prime_-} &=& \sqrt{1-\epsilon^2} \ket{J_z = -1/2} - 
\epsilon \ket{J_z = +3/2},
\label{hole1}
\eeqa
and 
\beqa
\ket{h_-} &=& \sqrt{1-\epsilon^2} \ket{J_z = -3/2} + 
\epsilon \ket{J_z = +1/2},\nonumber\\
\ket{h^\prime_+} &=& \sqrt{1-\epsilon^2} \ket{J_z = +1/2} - 
\epsilon \ket{J_z = -3/2},
\label{hole2}
\eeqa
where $\epsilon$ characterizes the degree of mixing.
$\ket{h_+}$ and $\ket{h_-}$ are degenerate states that are predominantly 
heavy-hole like, and which are split from the second degenerate pair, 
$\ket{h^\prime_+}$ and $\ket{h^\prime_-}$, which have predominantly 
light-hole character. The nature of mixing here is quite different 
to the type considered in Ref.~\onlinecite{calarco03}, where 
mixing between the $\ket{J_z=+3/2}$ and $\ket{J_z=+1/2}$ 
(or $\ket{J_z=-3/2}$ and $\ket{J_z=-1/2}$) states was assumed.

To estimate the hole mixing parameter
let us now consider a specific and very simple model: that of a hole bound in a parabolic 
potential in all three dimensions.~\cite{nazir05} This external potential 
is defined by $V(x, y, z) =(\w_x^2 x^2 +\w_y^2 y^2 
+ \w_z^2z^2)/\gamma_1$,~\cite{tsitsishvili98} where $\w_j$ is the 
frequency of the trapping potential in the 
$j=\{{\bf \hat{x},\hat{y},\hat{z}}\}$ direction.
We shall make the {\it axial approximation} -- i.e. 
we shall ignore terms which are not axially symmetric 
about the $z$-axis; this is a good approximation for 
GaAs~\cite{pedersen96} (corrections to this approximation are discussed in Ref.~\onlinecite{govorov05}). Then the Luttinger-Kohn Hamiltonian 
(Eq.~\ref{luttinger}) becomes, in the $\{\ket{J_z=+3/2}, 
\ket{J_z=-1/2}\}$ basis (or equivalently the $\{\ket{J_z=-3/2}, 
\ket{J_z=+1/2}\}$ basis):
\beq
\label{luttingerpara}
{\cal H}= \frac{1}{4}\left( \begin{array}{cc}
2 \w_T + \Delta E_h
& W \\
W &
2 \w_T-\Delta E_h
\end{array} \right),
\eeq
where, in order to simplify notation, we have defined: 
$\w_T\equiv\w_x+\w_y+\w_z$,
$\Delta E_h\equiv (\w_T-3\w_z)\frac{\gamma_2}{\gamma_1}$,
and $W\equiv\frac{\sqrt{3}(\gamma_2+\gamma_3)}{2}(\w_x-\w_y)$.

Assuming that the difference between the diagonal elements is much 
greater than the magnitude of the off-diagonal elements -- i.e.
$2 \Delta E_h \gg W$,
then $\epsilon\ll 1$ and is given by:
\beq
\epsilon \approx  \frac{W}{2 \Delta E_h}=\frac{\sqrt{3}\gamma_1(\gamma_2+\gamma_3)(\w_x-\w_y)}
{4\gamma_2(\w_T-3\w_z)}.
\eeq
As expected, the mixing is proportional to sub-band coupling and 
inversely proportional to the energy difference between the heavy and light
hole sub-bands. Taking typical values for the Luttinger 
parameters for GaAs ($\gamma_1 = 6.8$, $\gamma_2 = 2.1$ 
and $\gamma_3 = 2.9$), and estimating the different trapping frequencies 
for an anisotropic QD to be $\hbar\w_x=$10~meV, 
$\hbar\w_y=$11~meV and $\hbar\w_z=$45~meV, leads to 
$\epsilon \approx 0.1$. We shall use this value 
throughout the rest of the paper.

\subsection{Relevant Trion States}

We shall now assume that the hole states 
$\ket{h^\prime_+}$ and $\ket{h^\prime_-}$ are energetically 
distant enough from $\ket{h_+}$ and $\ket{h_-}$ that they may be 
ignored in our calculations of quantum dynamics 
(in the parabolic well model considered in the previous 
section this splitting is of order 11~meV). 
We are interested in using single excess spins in QDs to embody our qubit, 
and in exploiting spin-dependent exciton creation
to couple together spins in adjacent dots. A single spin becomes a 
trion state~\cite{trions} following the creation of an exciton, and by using 
Eqs.~\ref{hole1} and~\ref{hole2} we find that the trion eigenstates are:
\beqa
\label{trions1}
\ket{\Xp} = \ket{S_{\uparrow \downarrow}}\otimes\ket{h_+},\\
\label{trions2}
\ket{\Xm} = \ket{S_{\uparrow \downarrow}}\otimes\ket{h_-},
\eeqa
where $S_{\uparrow \downarrow}$ denotes two electrons in opposite spin states.

We emphasize that the states $\ket{h_+}$ and $\ket{h_-}$ are not eigenstates of the $J_z$ operator -- and this has profound consequences for both the coupling of charge carriers to the laser
field as well as the F\"orster transfer interaction.
We now specifically calculate the form of these two interactions.

\subsection{Interaction with a laser field}
\label{laser}

The QD-light interaction for a (classical) laser pulse of amplitude $E(t)$ and central frequency $\omega_L(t)$ impinging on a single QD may be expressed in the dipole approximation~\cite{haug90} through the following Hamiltonian operator:
\beq
\hat{{\cal H}}_{L}(t) = e E(t)\hat{{\bf r}} \cdot \hat{{\bf n}} \cos[\w_L(t) t]
\label{Hnorwa}
\eeq
where $\hat{\bf r}$ is the dipole operator, $\hat{{\bf n}}$ is the polarization vector of the light field, and $e$ is the electronic charge. The time dependence of $E$ and $\omega$ allows us later to introduce chirped laser pulse shapes: we assume that the time dependence is slow compared with the oscillation period of the laser.

We assume that the laser
pulse has a spectral width which is narrower the typical QD level spacing. If we choose the laser frequency to be close to resonance with the ground state exciton, we can
then consider an idealized model in which the dynamics is restricted only
to the two qubit states defined as 
\begin{eqnarray}
\ket{0} & = & \ket{-1/2_e} \nonumber \\
\ket{1} & = & \ket{ 1/2_e},
\end{eqnarray}
and to the trionic states $\ket{\Xp}$ and $\ket{\Xm}$.

By choosing the laser pulse to be directed along $z$ and
$\sigma^+$ circularly polarized, we find that $\hat{\bf r} \cdot \hat{{\bf n}} = (x +iy)$.
Eqs.~\ref{bloch1}-\ref{bloch4},~\ref{bloch7},~\ref{bloch8}, and~\ref{Hnorwa} then allow us to calculate the form of the 
interaction between the qubit electron spin and trion states. Let us first define the length 
\beq
l_i \equiv \int f_i(r) r^3 g(r) dr \ ,
\eeq
with $i\in \{lh, hh\}$. We also use the modified mixing parameter $\tilde{\epsilon}\equiv\epsilon \frac{l_{lh}}{l_{hh}\sqrt{3}}$
and the Rabi frequency
\beq
\Omega(t) \equiv \frac{2e E(t) l_{hh}}{\sqrt{6}}.
\eeq
Then we find, for a quantum dot labelled by $\kappa$:
\beq
{\cal H}_{\rm \sigma^+, \kappa} (t)= \Omega (t)\cos[\w_L(t) t]
\left( \ket{1}_\kappa\bra{\Xp} +\tilde{\epsilon}\ket{0}_\kappa\bra{\Xm} + H. c. \right).
\label{eq:lightham2}
\eeq

In the absence of mixing ($\tilde{\epsilon} = 0$) only the spin up qubit
state $\ket{1}$ is coupled to the laser field; the spin down state
$\ket{0}$ is completely decoupled. However, for a finite amount of hole mixing 
the spin-selectivity of trion excitations (required for the scheme of Ref.~\onlinecite{nazir04}) is no longer maintained
(see Fig.~\ref{mixing}) and we must therefore consider an alternative gating strategy.
\begin{figure}
\centerline{\hspace{0cm} \includegraphics[width=3.2in,height=2.1in]
{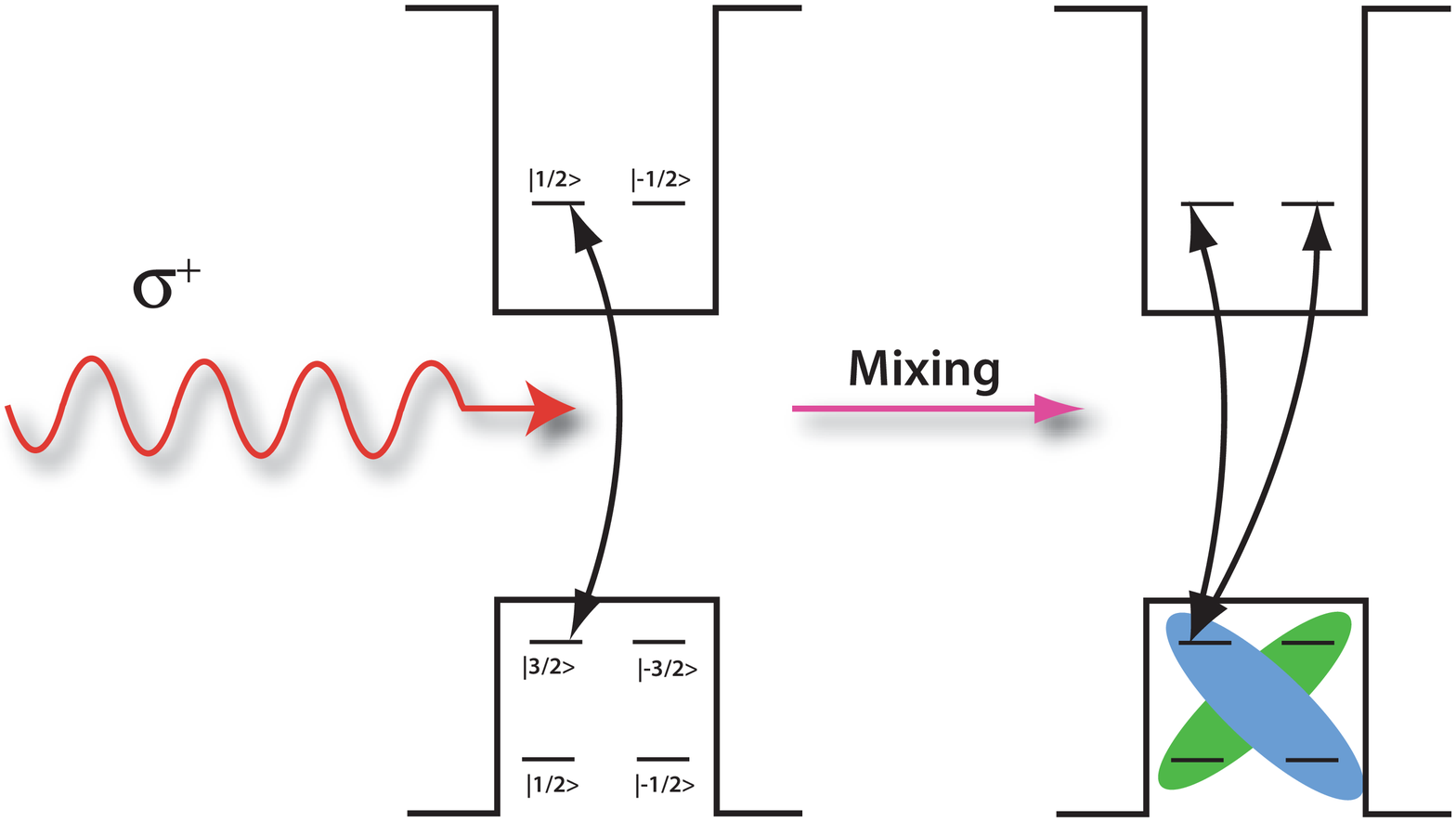}}
\caption{Hole states in the presence of hole sub-band mixing and
its effect on Pauli blocking. The arrows denote the allowed optical
inter-band transitions for an incoming $\sigma^{+}$ polarized laser 
pulse. The left part of the figure represents the situation where no hole
mixing is present and the right hand side shows the effects of mixing.
}
\label{mixing}
\end{figure}

\section{Coupled Quantum Dot Structure}
\label{coupledQD}

There are two principal interactions between trion states in adjacent quantum dots. The first of these is the direct Coulomb binding energy $V_{XX}$ between two trions, which leads to the biexcitonic shift.~\cite{Fausto}
The other interaction is the off-diagonal F\"orster coupling, which induces the transfer of electron hole pairs, through virtual photons, and we discuss this in the following sections.

\subsection{Angular momentum dependence of the F\"orster interaction}
\label{angmom}

The general form of the F\"orster coupling Hamiltonian in QDs was 
discussed in detail in Ref.~\onlinecite{lovett03}, where it was 
found that the magnitude of the interaction is given by
\beqa
V_F = &\frac{e^2}{4\pi\epsilon_0 \epsilon_r R^3} W_{\rm 1} W_{\rm 2}&
\left(\langle  {\bf r}_{\rm 1} \rangle \cdot \langle{\bf r}_{\rm 2} \rangle 
\right. \nonumber\\
&&\left. -\frac{3}{R^2} (\langle {\bf r}_{\rm1} \rangle \cdot {\bf R}) 
(\langle {\bf r}_{\rm2}\rangle \cdot {\bf R})\right),
\label{forster}
\eeqa
where ${\bf R}$ is the vector connecting the centers of the two QDs.
The term $\langle {\bf r}_i \rangle$ represents the interband 
expectation value of the atomic position operator for dot  $i$:
\beq
\label{bloch}
\langle {\bf r}_i \rangle = \int_{cell,~dot~i} U_e({\bf r}) {\bf r} 
U_h({\bf r}) d{\bf r}, 
\eeq
where the $U_e$ and $U_h$ represent the Bloch functions for 
electrons and holes within dot $i$. The $W_i$ represent the 
overlap of the envelope functions $\phi_e$ and $\phi_h$ for dot $i$:
\begin{equation}
W_i = \int_{space} \phi{\rm _e^i} ({\bf r}) \phi{\rm _h^i} ({\bf r}) 
d{\bf r}.
\label{overlap}
\end{equation}

An experiment would typically be performed on a pair of {\it vertically stacked} quantum dots, where the vector ${\bf R}$ lies along the growth direction $z$: by using this in Eq.~\ref{forster}, it is easy to show that the F\"orster interaction conserves the 
angular momentum of transferring excitons. However, the nature of the 
exciton state before and after the transfer affects the magnitude $V_F$. 
By substituting the Bloch functions of Eqs.~\ref{bloch1} to~\ref{bloch6} into 
Eq.~\ref{forster} and using the relations of Eqs.~\ref{rel1} to \ref{rel2}, 
we can obtain the strength of the F\"orster coupling for excitons which 
are composed of electrons and holes of varying angular momentum. 
We defined
\beq
\label{lambda}
M_{ij} = \frac{e^2}{12\pi\epsilon_0 \epsilon_r R^3} 
W_{\rm 1} W_{\rm 2} l_i l_j,
\eeq
for $i,j\in \{lh, hh\}$. Table~I then shows the matrix element for all 
transitions which are induced by the F\"orster interaction.

\begin{table*}
\label{dipoles}
\begin{center}
\begin{tabular}{|c|c|c|c|} \hline
 State 1 ($J_{z,h}$, $J_{z,e}$) & State 2 ($J_{z,h}$, $J_{z,e}$) & 
Matrix Element & Net $J_z$ \\ \hline
$-\frac{3}{2}$; $\frac{1}{2}$ & $-\frac{3}{2}$; $\frac{1}{2}$ 
& $M_{hh,hh}$ & $-1 \leftrightarrow -1$ \\
$\frac{3}{2}$; $-\frac{1}{2}$ & $\frac{3}{2}$; $-\frac{1}{2}$ 
& $M_{hh,hh}$ & $+1 \leftrightarrow +1$ \\
$-\frac{1}{2}$; $\frac{1}{2}$ & $-\frac{1}{2}$; $\frac{1}{2}$ 
& $-4M_{lh,lh}/3$ & $0 \leftrightarrow 0$ \\
$\frac{1}{2}$; $-\frac{1}{2}$ & $\frac{1}{2}$; $-\frac{1}{2}$ 
& $-4M_{lh,lh}/3$ & $0 \leftrightarrow 0$ \\
$-\frac{1}{2}$; $\frac{1}{2}$ & $\frac{1}{2}$
; -$\frac{1}{2}$ & $4M_{lh,lh}/3$ & $0 \leftrightarrow 0$ \\
$-\frac{1}{2}$; -$\frac{1}{2}$ & -$\frac{1}{2}$; -$\frac{1}{2}$ 
& $M_{lh,lh}/3$ & $-1 \leftrightarrow -1$ \\
$\frac{1}{2}$; $\frac{1}{2}$ & $\frac{1}{2}$; $\frac{1}{2}$ 
& $M_{lh,lh}/3$ & $+1 \leftrightarrow +1$ \\
$-\frac{3}{2}$; $\frac{1}{2}$ & $-\frac{1}{2}$; -$\frac{1}{2}
$ & $M_{lh,hh}/\sqrt{3}$ & $-1 \leftrightarrow -1$ \\
$\frac{3}{2}$; $-\frac{1}{2}$ & $\frac{1}{2}$; $\frac{1}{2}$ 
& $M_{lh,hh}/\sqrt{3}$ & $+1 \leftrightarrow +1$ \\\hline
\end{tabular}
\caption{Relative size of the matrix element for exciton 
states coupled by the F\"orster interaction. $M$ is defined in 
Eq.~\ref{lambda}.}
\end{center}
\end{table*}

\subsection{F\"orster interaction for coupled trions}
We next derive the form of the 
F\"orster transfer operator $\hat{T}$ for two coupled trions. The most general definition of $\hat{T}$ is:
\beq
\hat{T} = \sum_{i,j,k,l} V_{F}^{i,j,k,l}\left(\ket{J_{z,e}^i, J_{z,h}^j;vac}\bra{vac;J_{z,e}^k, J_{z,h}^l} +H. c. \right)
\eeq
where $V_F^{i,j,k,l}$ is the size of the F\"orster matrix element which connects an electron and hole on dot 2 (whose angular momentum states are labelled with indices $k$ and $l$) to an electron and hole and dot 1 (whose angular momentum states are labelled with indices $i$ and $j$). The only non-zero values of $V^F_{i,j,k,l}$ are given in Table 1, and we use that table to work out the effect of $\hat{T}$ on states in our 
two-dot system. For example, we have that:
\beqa\nonumber
\hat{T}\ket{1{\Xp}}&=&\hat{T} \ket{+\frac{1}{2}_e;+\frac{3}{2}_h, S_{\uparrow \downarrow}}
+\epsilon\ket{+\frac{1}{2}_e; -\frac{1}{2}_h, S_{\uparrow \downarrow}}\\
&=&M_{hh,hh}\ket{+\frac{3}{2}_h, S_{\uparrow \downarrow}; +\frac{1}{2}_e} 
\nonumber\\ 
&&+ \frac{\epsilon M_{lh,lh}}{3}\ket{-\frac{1}{2}_h, S_{\uparrow \downarrow}; 
+\frac{1}{2}_e} \nonumber\\&&+\frac{4\epsilon M_{lh,lh}}{3}
\ket{+\frac{1}{2}_h, S_{\uparrow \downarrow}; -\frac{1}{2}_e}.
\eeqa
We can now see that
\beqa
\bra{\Xp1}\hat{T}\ket{1\Xp}& 
=& M_{hh,hh}+\frac{\epsilon^2M_{lh,lh}}{3},\\
\bra{\Xm 0}\hat{T}\ket{1\Xp}&=& \frac{4\epsilon^2M_{lh,lh}}{3}.
\eeqa

Similar calculations allow us to find all of the F\"orster coupling terms, 
which may be expressed by the following Hamiltonian, correct to first order 
in $\epsilon$:
\beqa
\label{forsimp}
{\cal H}_F &=&  M_{hh, hh} (\ket{0{\Xm}}\bra{{\Xm}0} + \ket{1{\Xp}}
\bra{{\Xp}1})\nonumber\\
&+&\frac{2M_{hh,lh}\epsilon}{\sqrt{3}} \left(\ket{1 \Xm}\bra{\Xp0} +
\ket{\Xm 1}\bra{0\Xp}\right) + H. c.\nonumber\\
\eeqa

\subsection{Full Hamiltonian}
The total Hamiltonian of the two QD system in the presence of laser excitation may now be written as:
\beqa
{\cal H}_T (t) & = &  \sum_{\kappa=a,b} \left[\delta\ket{1}_{\kappa}\bra{1} 
+ \w_{X_\kappa} \hat{P}_{X_\kappa} +{\cal H}_{\sigma^+, \kappa} (t)\right ]+ {\cal H}_F
\nonumber \\
&+& \sum_{\nu, \mu \in \{\Xp,\Xm\}} V_{XX} (\ket{\nu\mu}\bra{\nu\mu}).
\eeqa
The state $\ket{00}$ sets the zero of 
energy -- and then the first term describes the Zeeman energy 
splitting $\delta$ of the spin qubits; the second term represents 
the trion creation energy $\w_{X_\kappa}$ where $\hat{P}_{X_\kappa} \equiv
 \ket{\Xp}_\kappa \bra{\Xp} + \ket{\Xm}_\kappa \bra{\Xm} $ is the projection 
operator onto the single trion state located in the QD labeled by
$\kappa$; the next two terms are defined by Eqs.~\ref{eq:lightham2} 
and~\ref{forsimp}, respectively, and the last term describes the static
dipole-dipole binding energy between trions 
$V_{XX}$.~\cite{lovett03, calarco03}  

We now move to a frame which is rotating at the laser frequency 
$\w_L$ for both spin-trion transitions, and make the Rotating 
Wave Approximation (RWA). We then find that:
\beqa
\label{finalham}
{\cal H}_T (t)& = &  \sum_{\kappa=a,b} \left [
\delta  \ket{1}_\kappa \bra{1} 
+ \Delta_{\kappa}(t)\hat{P}_{X_\kappa} 
+{\cal H}_{\sigma^+, \kappa}^\prime (t)\right ]+  {\cal H}_F
\nonumber\\
&+& \sum_{\nu, \mu \in \{\Xp,\Xm\}} V_{XX} \ket{\nu\mu}\bra{\nu\mu},
\eeqa
where $\w_{X_\kappa}$ has now been replaced by the time
dependent detuning of the laser 
from the spin to trion transition energy
$\Delta_{\kappa}(t) \equiv \omega_{X_\kappa}-\omega_L(t)$.
The charge-laser field coupling is now given by
\beq
\label{lighthamprime}
{\cal H}_{\rm \sigma^+, \kappa}^\prime (t)=  \frac{\Omega(t)}{2} 
\left( \ket{1}_\kappa\bra{\Xp} +\tilde{\epsilon}
\ket{0}_\kappa\bra{\Xm} + H. c. \right).
\eeq

\begin{figure}
\centerline{\hspace{0cm} 
\includegraphics[width=4in,height=4in]{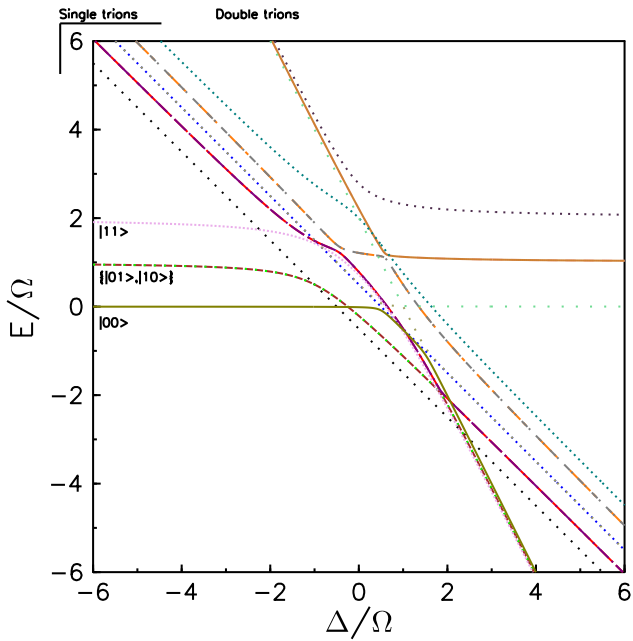}}
\caption{Eigenstate energy spectrum. The parameters are set as follows: 
$\delta/\Omega =1$,
$M_{hh,hh}/\Omega  = M_{lh,hh}/\Omega=0.5$, $V_{XX}/\Omega = 2$, $\epsilon = 0.1$.}
\label{eigenstates}
\end{figure}

The Hamiltonian, Eq.~\ref{finalham}, spans a 16-dimensional Hilbert space. 
It is composed of the four computational basis states 
($\ket{00}$, $\ket{01}$, $\ket{10}$ and $\ket{11}$), eight single 
trion states, and four double trion states. In order to get a little 
more insight into the behavior of the system, we plot its 
eigenenergies as a function of the ratio $\Delta/\Omega$ in 
Fig.~\ref{eigenstates}. There are three main groups of curves 
that are well separated away from $\Delta/\Omega = 0$ (i.e. when $|\Delta/\Omega|\gg1$); these groups are simply the four computational 
basis states, eight single trion states and four double trion states. As 
$\Delta/\Omega$ approaches zero (i.e. when the laser becomes resonant 
with the spin to trion transition energies), the eigenstates become 
superpositions involving different numbers of trions, and many 
anticrossings can be seen in the eigenstate spectrum. 

\label{sec:mixing}

\section{Adiabatic Quantum Gates}
\label{gates}

In this section, we shall describe how an adiabatic change in the ratio $\Delta/\Omega$ can allow us to follow the eigenstate curves. Not all of the computational basis states mix in the same way with trions when $\Delta/\Omega$ is small. Following Ref.~\onlinecite{calarco03}, it is possible to use chirped laser pulses to slowly vary $\Delta/\Omega$, which causes a non-trivial two qubit operation via trion state anticrossings. 

\subsection{Chirped pulses}
\label{chirped}
Let us consider such a chirped laser pulse, where the detuning varies in time as 
follows:
\beq
\Delta(t) = -\Delta_0 \left( 1 -\frac{1}{2}e^{-(t/\tau_\Delta)^2} \right).
\eeq
$\Delta_0$ represents the maximum detuning and $\tau_\Delta$ is a 
parameter characterizing the time variation of the detuning.

The time dependence of the laser intensity is assumed to cause the Rabi 
frequency to take the following form:
\beq
\Omega (t) = \Omega_0 e^{-(t/\tau_\Omega)^2} ,
\eeq
where, similarly, $\Omega_0$ represents the maximum Rabi frequency 
and $\tau_\Omega$ is a parameter characterizing its time variation. For a review of experimental methods for pulse shaping see Ref.~\onlinecite{warren93}.

We calculate the quantum dynamics caused by our Hamiltonian with these 
time-varying laser pulses by using a numerical Schr\"odinger equation 
solver. If the adiabatic approximation holds true, population will 
return to the initial state at the end of the operation (non-adiabatic
corrections limit the gate fidelity and are described below). However, the phase 
accumulated during the gate varies depending on the initial state. 
We characterize the gate by looking at the relative phase $\theta$ 
gained when the pulse is applied to each of the four computational 
basis states in turn:~\cite{calarco03}
\beq
\theta \equiv \phi_{00} - \phi_{01} - \phi_{10} +\phi_{11}
\label{theta}
\eeq
where $\phi_{n}$ is the phase change of $\ket{n}$ during the gate operation. 
The relative phase $\theta$ is 
the part of the phase which is invariant under single qubit 
operations (see Ref.~\onlinecite{vager05} for a detailed discussion). We can see this by considering the effect of the following gates. First,
\beq
U_1= \left( \begin{array}{cc}
e^{-i\phi_{00}} & 0  \\
0 & e^{-i\phi_{10}} \\  
\end{array} \right)
\eeq
is performed on qubit 1 (in the $\ket{0}$, $\ket{1}$ basis), and then
\beq
U_2= \left( \begin{array}{cc}
1 & 0  \\
0 & e^{i(\phi_{00}-\phi_{10})} \\  
\end{array} \right)
\eeq
is performed on qubit 2. These two single qubit gate operations remove any phase picked up on the states $\ket{00}$, $\ket{01}$ and $\ket{10}$, with $\ket{11}$ undergoing a net phase change of $\theta$. In the basis $\ket{00}$, $\ket{01}$, $\ket{10}$, $\ket{11}$, a CPHASE gate is
\beq
U_{CPHASE}= \left( \begin{array}{cccc}
1 & 0  & 0 & 0\\
0 & 1 & 0 & 0 \\  
0 & 0 & 1 & 0 \\  
0 & 0 & 0 & -1   
\end{array} \right).
\eeq
This can be constructed from any gate operation in which $\theta=\pi$, together with appropriate single qubit gates.~\cite{nazir04} 

\begin{figure}[h]
\centerline{\hspace{0cm} 
\includegraphics[width=3.5in,height=3.5in]{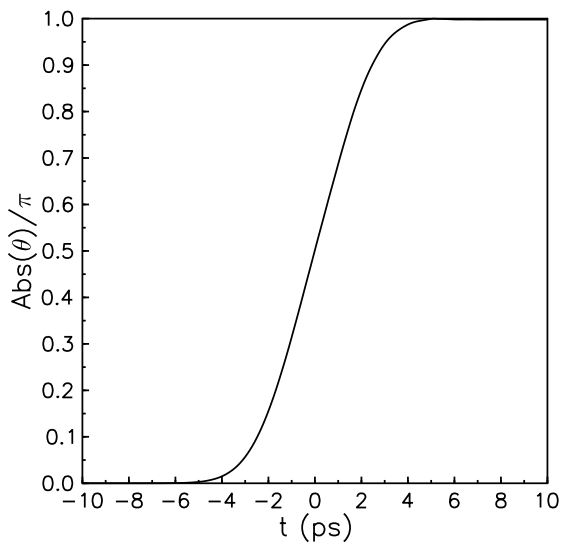}}
\caption{Variation of $\theta$ (Eq.~\ref{theta}) as a function of time. 
The parameters are set as follows: 
$\delta=1$~meV,
$M_{hh,hh}=M_{lh,hh} =0.5$~meV,
$V_{XX}=2$~meV,
$\epsilon=0.1$,
$\tau_\Omega=3.55$~ps, $\tau_\Delta=2.55$~ps, 
$\Delta_0=4.5$~meV and $\Omega_0=8$~meV. }
\label{thetaVXX}
\end{figure}

\begin{figure}[h]
\centerline{\hspace{0cm} 
\includegraphics[width=3.5in,height=3.5in]{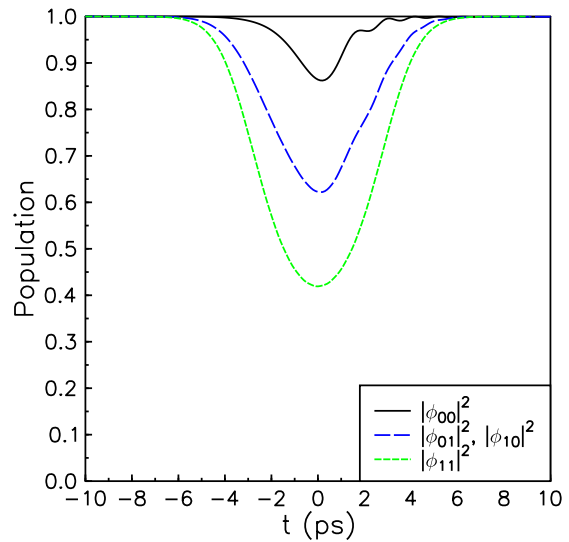}}
\caption{Variation of the initial state 
population for each computational basis state during a gate operation. 
The parameters are the same as in Fig.~\ref{thetaVXX}.}
\label{popsVXX}
\end{figure}

A strength of our approach is that a phase gate can be performed whether the interdot coupling takes the diagonal form (biexcitonic interaction) or an off-diagonal form (F\"orster transfer).
We can illustrate this by performing numerical simulations in two limits.
First, we consider the case where 
$V_{XX} > M_{hh,hh} =  M_{lh,hh} $.
In this case the required
phase is given predominantly through the biexcitonic coupling.
If we take the same parameters as used for Fig.~\ref{eigenstates}, with 
chirped pulses characterized by $\tau_\Omega=3.55$~ps,  
$\tau_\Delta =2.55$~ps, $\Delta_0 =4.5$~meV and $\Omega_0 =8$~meV, 
then we can indeed obtain $\theta=\pi$. This is shown in Fig.~\ref{thetaVXX}, 
which displays the time dependence of $\theta$.
The population of each computational basis state (following initialization 
into that state) is displayed in Fig.~\ref{popsVXX}; the
chirped pulse moves some of the population out of each state (and into trion states)
and then back 
again. The effect of the pulse is the same for $\ket{01}$ and $\ket{10}$, 
as would be expected on further inspection of our Hamiltonian. However, 
there is a distinct difference between the behavior of these 
two states and that of the states $\ket{00}$ and $\ket{11}$ -- and it is 
this difference which allows the relative entangling phase 
$\theta$ to be picked up. 

\begin{figure}[h]
\centerline{\hspace{0cm} \includegraphics[width=3.5in,height=3.5in]
{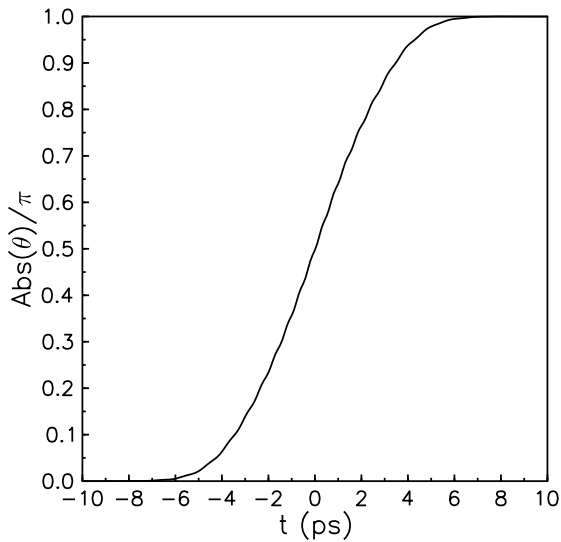}}
\caption{Variation of $\theta$ (Eq.~\ref{theta}) as a function of time. 
The parameters are set as follows: 
$\delta$ = 1~meV,
$M_{hh, hh} =  M_{lh,hh} =$ 0.5~meV,
$V_{XX} =$ 0,
$\Omega_0 =$ 8~meV,
$\tau_\Omega =$ 4.2~ps,
$\tau_\Delta = $3~ps,
$\Delta_0 = $3~meV and
$\epsilon = 0.1$.}
\label{thetaVXX0}
\end{figure}

\begin{figure}
\centerline{\hspace{0cm} \includegraphics[width=3.5in,height=3.5in]
{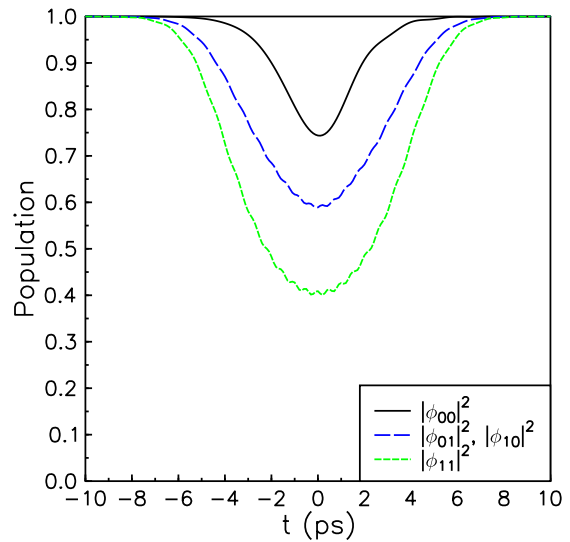}}
\caption{Variation of the initial state population for each computational 
basis state during a gate operation. The parameters are the same as in 
Fig.~\ref{thetaVXX0}.}
\label{popsVXX0}
\end{figure}

We now move to the second case, where $V_{XX}=0$.  
We use the parameters
$\delta=1$~meV,
$M_{hh, hh} = M_{lh,hh}=0.5$~meV,
$V_{XX} =0$,
$\Omega_0=8$~meV,
$\tau_\Omega=4.2$~ps,
$\tau_\Delta=3$~ps,
$\Delta_0=3$~meV and
$\epsilon=0.1$; in Figs.~\ref{thetaVXX0} 
and~\ref{popsVXX0} we show the time variation of 
$\theta$ and the variation of basis state populations.
We can see that it is possible to pick 
up an entangling phase ($\theta=\pi$) in this 
case too, and the effect now relies solely on the 
F\"orster interaction.

\subsection{Gate fidelity}
\label{fidelity}

The gate fidelity is limited by a number of factors including: corrections to our 
effective four level QD model, spontaneous emission from excited states, 
coupling of the charge carriers to
the underlying lattice which is responsible for coupling to phonons, and the 
possibility of non-adiabatic transitions due to the finite gate operation time.

Corrections to the effective four level QD model can be 
approximated as exponentially small in the ratio of the spectral width of
laser pulse to the energy level spacing. As we discussed in Section~\ref{laser}, this ratio is much smaller than unity, and the reduction in the fidelity
due to excitation to higher energy levels of the QDs can be safely
ignored.

It was shown in Ref.~\onlinecite{calarco03} that, as well as avoiding 
difficulties due to hole mixing, the adiabatic gating scheme also avoids further unwanted transitions related to phonon decoherence.
However, since the gate is operated in a 
finite time, which one wishes to minimize, there are 
necessarily non-adiabatic transitions between the laser dressed states.
Such non-adiabatic transitions are described by the Landau-Zener (LZ)
theory.~\cite{LZ} If we assume a constant rate of change for the detuning, i.e.
$\Delta(t)=\dot{\Delta}t$, the condition for adiabaticity is given by
$\Omega^2/\dot{\Delta} \gg 1$, where $\Omega$ is the energy separation of the levels at closest approach. The probability for an unwanted transition by
$P=\exp{(-\pi \Omega^2/4\dot{\Delta})}$. There is a simple physical interpretation of this~\cite{calarco03}: 
$\tau\sim\Omega/\dot{\Delta}$ is the characteristic time of sweep
through the resonance, and so the adiabatic condition is 
naturally $\Omega\tau \gg 1$. In the non-linearized version of LZ theory the
dependence of the unwanted transition is still an exponentially 
small function of both the reciprocal of the energy difference between the two levels 
($\Delta E_{\alpha \beta} \sim 1/\Omega_{\alpha \beta}$ where
$\alpha, \beta $ are two general QD states)
and the characteristic sweep time
$\tau$. The proposed adiabatic gate scheme is based on the fact
that the different avoided crossings between the laser
dressed states can be easily distinguished, i.e. different avoided
crossings occur for different values of detunings $\Delta$ and energies
$E$ (see Fig.~\ref{eigenstates}). For the adiabatic gate scheme in the 
limit $M_{hh,hh}=M_{lh,hh}<V_{XX}$ the relevant Rabi frequency is the
coupling between
the states, $\ket{00}$, and $\ket{11}$ when $\Delta\approx 0$, which is the 
biexcitonic shift $V_{XX}$.
Therefore one can estimate the probability of unwanted transitions 
to be around $10^{-6}$ for a typical sweep time of $\tau\sim4$~ps.
For the case in which the gate is based solely on the 
F\"orster transfer interaction $M_{hh, hh} = M_{lh,hh}=0.5$~meV
the fidelity is lower and the probability of unwanted transitions for
the same typical sweep time is of the order of a few percent.
In order to get a better gate fidelity for
the gate based on F\"orster interaction the typical sweep time $\tau$ should
be made longer. 

The adiabatic gate procedure may also reduce the effects of phonon decoherence.~\cite{calarco03} If the gate
is operated in the bi-excitonic mode the probability of unwanted 
transitions and decrease in pure dephasing decoherence effects~\cite{kuhn02,pazy02,Jacak02} is expected to be of the form
\beq
P\sim \frac{J(\omega_{m})}{\Omega }\exp(-\lambda \Omega \tau )\ ,
\label{eq:phonon}
\eeq
where $\lambda$ is a positive constant of order unity and 
$J(\omega_{m})$ is the spectral function that describes the coupling of phonons to our system 
evaluated at some high-frequency cut-off $\omega _{m}$ that is
imposed by the speed of the frequency detuning sweep. 

The typical time scale for spin dephasing is of the order of $\mu$s and can be made even longer
by optical pumping of nuclear spins.~\cite{Bracker} Polarizing the
nuclear spins will increase spin coherence since the main 
mechanism for spin dephasing is the coupling of the electronic spin
to the nuclear spins.~\cite{marcus05} Thus the ps time scale for the adiabatic gate 
does not constitute too strong a restriction.

\section{State Preparation, Measurement and Scalability}
\label{meas}
Though our paper focuses on an adiabatic two qubit gate scheme which allows
one to resolve the difficulties arising due to hole mixing, we would
 like to briefly comment on the possibility for optically fulfilling the
other requirements for a quantum information implementation scheme,
i.e., initial state preparation, measurement, and scalability. 

The essential test for any implementation scheme is of course provided by
experiment. Recently there has been tremendous experimental effort and 
success in validating the essential stages needed for solid state quantum
computation implementation schemes employing optically driven charged QDs.
In a new experiment Gurudev Dutt {\it et. al.}
have demonstrated that a coherent optical field can produce 
coherent electronic spin states in QDs, demonstrating that these
electronic spin states have life times much longer than the exciton 
coherence times.~\cite{GurudevDutt05}
Other experiments
have shown how the spin state of the resident electron in a
self-assembled InAs-GaAs QD can be written and read using
circularly polarized optical pumping.~\cite{Cortez,Bracker}
Methods for an optical read-out mechanism of the spin of 
an electron confined to a QD have also been theoretically
suggested,~\cite{Shabaev03,pazy04,Gywat04} and some experimental
work towards this goal has already been performed for colloidal semiconductor
QDs.~\cite{Lifshitz}
Scalability is a potential problem, since spatial selectively of individual qubits is not possible in our system. This is because
the optical wave length of the exciting laser pulse is 
much larger then the inter-dot distance needed to couple our qubits. 
To optically resolve different QDs we would need to resort to energy-selective 
addressing methods within different QD clusters whose size can be 
controlled.~\cite{Rich} Inside each QD cluster one could spectrally
differentiate a QD by applying a gate potential and inducing a Stark shift of the exciton levels.~\cite{nazir05} Alternatively, globally applied pulses could be used within a cellular-automaton scheme.~\cite{Fausto, benjamin00} One other possibility is that small scale processors based on our scheme could be joined together by using the exciton coupling to single photons. Linear optics techniques could then be used to entangle the states of two small processors.~\cite{barrett04, beige04}

\section{Summary}

To summarize, we have derived the Hamiltonian for a pair of spin qubits in adjacent QDs, 
which are coupled by trion states. Our analysis shows that it is possible to
perform a non-trivial two qubit gate in this system by using chirped 
laser pulses, even in the presence of hole mixing, and for two different 
types of interaction. In Ref.~\onlinecite{calarco03} it was shown that a 
dipole-dipole interaction can be used to mediate an adiabatic gate of 
this type. Our extension of that approach to include F\"orster processes generalizes this to cover all significant excitonic interactions in a coupled dot system.

In contrast to our previous work,~\cite{nazir04} the gate proposed here 
could be performed in many different materials, since we have shown that 
it is generally valid for all significant excitonic interactions 
and for varying degrees of hole mixing. We therefore believe that 
demonstration experiments could be performed in the near future.

\label{summary}

\section{Acknowledgments}
This research is part of the QIP IRC www.qipirc.org (GR/S82176/01) and is supported through the Foresight LINK Award Nanoelectronics at the Quantum Edge www.nanotech.org by EPSRC (GR/R660029/01) and Hitachi Europe Ltd. GADB thanks EPSRC for a Professorial Research Fellowship (GR/S15808/01).
SDB is supported by the EU projects Nanomagiqc and Ramboq.
We thank S.~C.~Benjamin
for useful and stimulating discussions. We are indebted 
to R.~G.~Beausoleil, who provided numerical simulation code.

\pagebreak


\end{document}